\documentclass[twoside,twocolumn,9pt]{article}
\usepackage{extsizes}
\usepackage[super,sort&compress,comma]{natbib} 
\usepackage[version=3]{mhchem}
\usepackage[left=1.5cm, right=1.5cm, top=1.785cm, bottom=2.0cm]{geometry}
\usepackage{balance}
\usepackage{times,mathptmx}
\usepackage{sectsty}
\usepackage{graphicx} 
\usepackage{lastpage}
\usepackage[format=plain,justification=justified,singlelinecheck=false,font={stretch=1.125,small,sf},labelfont=bf,labelsep=space]{caption}
\usepackage{float}
\usepackage{fancyhdr}
\usepackage{fnpos}
\usepackage[english]{babel}
\addto{\captionsenglish}{%
  
}
\usepackage{array}
\usepackage{droidsans}
\usepackage{charter}
\usepackage[T1]{fontenc}
\usepackage[usenames,dvipsnames]{xcolor}
\usepackage{setspace}
\usepackage[compact]{titlesec}
\usepackage{hyperref}

\definecolor{cream}{RGB}{222,217,201}

\begin{document}

\pagestyle{fancy}
\thispagestyle{plain}
%

\makeFNbottom
\makeatletter
\renewcommand\LARGE{\@setfontsize\LARGE{15pt}{17}}
\renewcommand\Large{\@setfontsize\Large{12pt}{14}}
\renewcommand\large{\@setfontsize\large{10pt}{12}}
\renewcommand\footnotesize{\@setfontsize\footnotesize{7pt}{10}}
\makeatother

\renewcommand{\thefootnote}{\fnsymbol{footnote}}
\renewcommand\footnoterule{\vspace*{1pt}%
\color{cream}\hrule width 3.5in height 0.4pt \color{black}\vspace*{5pt}} 
\setcounter{secnumdepth}{5}

\makeatletter 
\renewcommand\@biblabel[1]{#1}            
\renewcommand\@makefntext[1]%
{\noindent\makebox[0pt][r]{\@thefnmark\,}#1}
\makeatother 
\renewcommand{\figurename}{\small{Fig.}~}
\sectionfont{\sffamily\Large}
\subsectionfont{\normalsize}
\subsubsectionfont{\bf}
\setstretch{1.125} 
\setlength{\skip\footins}{0.8cm}
\setlength{\footnotesep}{0.25cm}
\setlength{\jot}{10pt}
\titlespacing*{\section}{0pt}{4pt}{4pt}
\titlespacing*{\subsection}{0pt}{15pt}{1pt}

\fancyfoot{}
\fancyfoot[LO,RE]{\vspace{-7.1pt}\includegraphics[height=9pt]{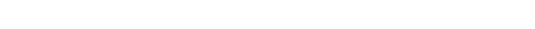}}
\fancyfoot[CO]{\vspace{-7.1pt}\hspace{13.2cm}\includegraphics{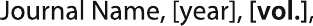}}
\fancyfoot[CE]{\vspace{-7.2pt}\hspace{-14.2cm}\includegraphics{head_foot/RF}}
\fancyfoot[RO]{\footnotesize{\sffamily{1--\pageref{LastPage} ~\textbar  \hspace{2pt}\thepage}}}
\fancyfoot[LE]{\footnotesize{\sffamily{\thepage~\textbar\hspace{3.45cm} 1--\pageref{LastPage}}}}
\fancyhead{}
\renewcommand{\headrulewidth}{0pt} 
\renewcommand{\footrulewidth}{0pt}
\setlength{\arrayrulewidth}{1pt}
\setlength{\columnsep}{6.5mm}
\setlength\bibsep{1pt}

\makeatletter 
\newlength{\figrulesep} 
\setlength{\figrulesep}{0.5\textfloatsep} 

\newcommand{\topfigrule}{\vspace*{-1pt}%
\noindent{\color{cream}\rule[-\figrulesep]{\columnwidth}{1.5pt}} }

\newcommand{\botfigrule}{\vspace*{-2pt}%
\noindent{\color{cream}\rule[\figrulesep]{\columnwidth}{1.5pt}} }

\newcommand{\dblfigrule}{\vspace*{-1pt}%
\noindent{\color{cream}\rule[-\figrulesep]{\textwidth}{1.5pt}} }

\makeatother

\twocolumn[
  \begin{@twocolumnfalse}
\vspace{3cm}
\sffamily
\begin{tabular}{m{4.5cm} p{13.5cm} }

\includegraphics{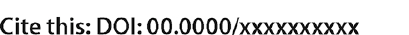} & \noindent\LARGE{\textbf{Theory of the splay nematic phase: Single vs.\ double splay}} \\
\vspace{0.3cm} & \vspace{0.3cm} \\

 & \noindent\large{Michely P. Rosseto\textit{$^{a}$} and Jonathan V. Selinger\textit{$^{b}$}} \\

\includegraphics{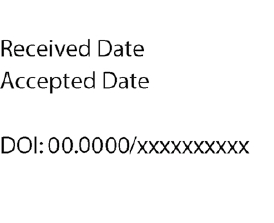} & \noindent\normalsize{Recent experiments have reported a novel splay nematic phase, which has alternating
domains of positive and negative splay. To model this phase, previous studies have considered a 1D splay modulation of the director field, accompanied by a 1D modulation of polar order. When the flexoelectric coupling between splay and polar order becomes sufficiently strong, the uniform nematic state becomes unstable to the formation of a modulated phase. Here, we re-examine this theory in terms of a new approach to liquid crystal elasticity, which shows that pure splay deformation is double splay rather than planar single splay. Following that reasoning, we propose a structure with a 2D splay modulation of the director field, accompanied by a 2D modulation of polar order, and show that the 2D structure generally has a lower free energy than the 1D structure.} \\

\end{tabular}

 \end{@twocolumnfalse} \vspace{0.6cm}

  ]

\renewcommand*\rmdefault{bch}\normalfont\upshape
\rmfamily
\section*{}
\vspace{-1cm}


\footnotetext{\textit{$^{a}$~Departamento de F\'isica, Universidade Estadual de Maring\'a, Maring\'a, Paran\'a 5790-87020-900, Brazil}} 
\footnotetext{\textit{$^{b}$~Department of Physics, Advanced Materials and Liquid Crystal Institute, Kent State University, Kent, Ohio 44242, USA}}





\section{Introduction}
\label{introduction}

Liquid crystals often exhibit modulated phases, which are induced by different types of molecular asymmetry.  The most common type of asymmetry is chirality.  When molecules are chiral, they tend to pack with a spontaneous twist.  This twist leads to the formation of cholesteric phases or blue phases.

Another type of asymmetry is a bent molecular shape, as occurs in dimers, trimers, and bent-core liquid crystals.~\cite{jakli2018}  In the nematic phase of bent molecules, the bend flexoelectric effect is enhanced\cite{meyer1969,prost1977,harden2006} and the bend elastic constant is reduced, because bend deformation is compatible with the molecular shape.  As the temperature decreases, the bend elastic constant decreases further, and then the system has a transition from the uniform nematic phase into a modulated phase with spontaneous bend.  Because it is impossible to fill space with pure uniform bend, the phase must have a more complex structure that includes another deformation mode.  Normally, the system forms a twist-bend nematic ($N_{TB}$) phase, which has a heliconical structure with twist as well as bend.  This phase has been investigated through many theoretical~\cite{meyer1976,dozov2001,memmer2002,shamid2013,meyer2013,virga2014,barbero2015,vaupotic2016,meyer2016} and experimental~\cite{chen2013,borshch2013,adlem2013,chen2014,meyer2015,gorecka2015,wang2016} studies.  An alternative theoretical possibility is a splay-bend nematic ($N_{SB}$) phase, which has a planar structure with splay as well as bend.~\cite{meyer1976,dozov2001,shamid2013}

Considering that a bent molecular shape leads to spontaneous bend, one might ask whether a splayed or pear-like shape leads to spontaneous splay.  This issue was considered in a theoretical paper,~\cite{Dhakal2010} which showed that a pear-like asymmetry gives an enhancement of the splay flexoelectric effect, which increases further as the temperature decreases.  That paper briefly mentioned simulations showing a polar phase with regions of splay separated by domain walls, but suggested that this structure was unlikely to occur in experiments.

In the last two years, experiments have actually reported a modulated phase induced by spontaneous splay.\cite{mertelj2018,mandle2019,connor2020,sebastian2020}  In these experiments, the splay elastic constant is reduced, because splay deformation is compatible with molecular shape.  As the temperature decreases, the splay elastic constant decreases further, and then the system has a transition from the uniform nematic phase into a modulated phase, which has been called the splay nematic ($N_S$) phase.

\begin{figure*}
\centering
(a)\includegraphics[width=0.68\linewidth]{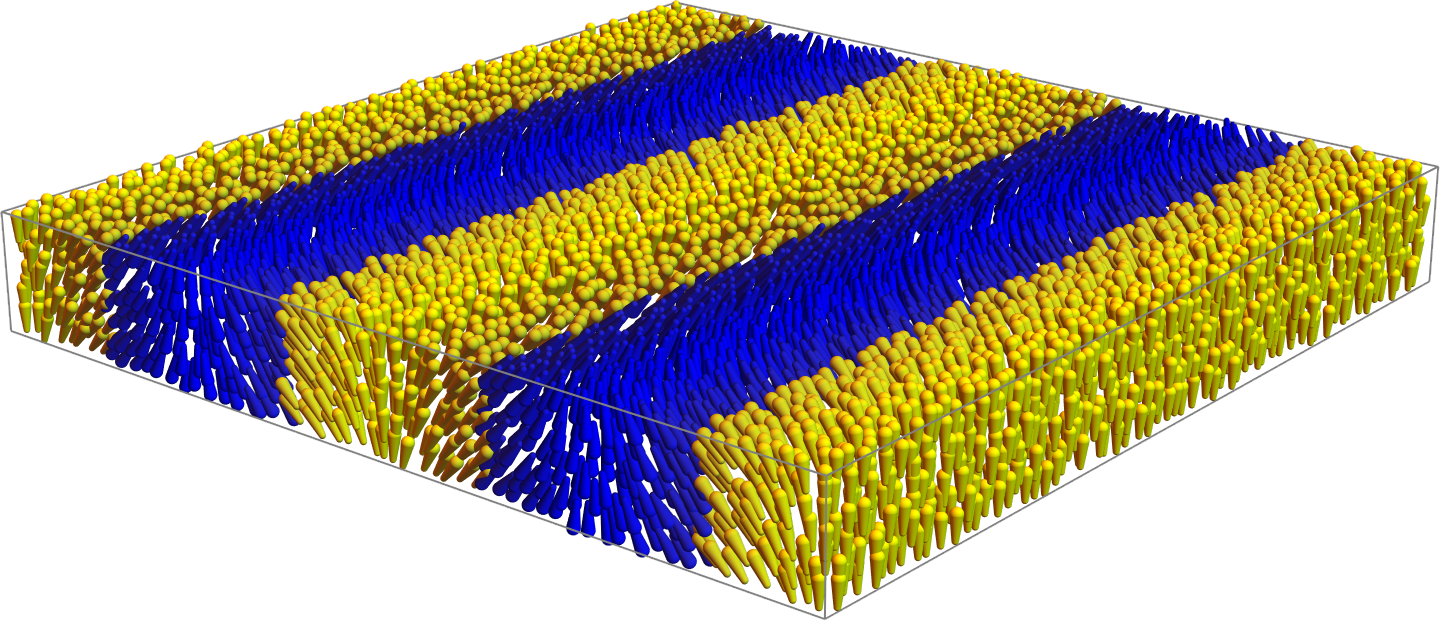}
(b)\includegraphics[width=0.68\linewidth]{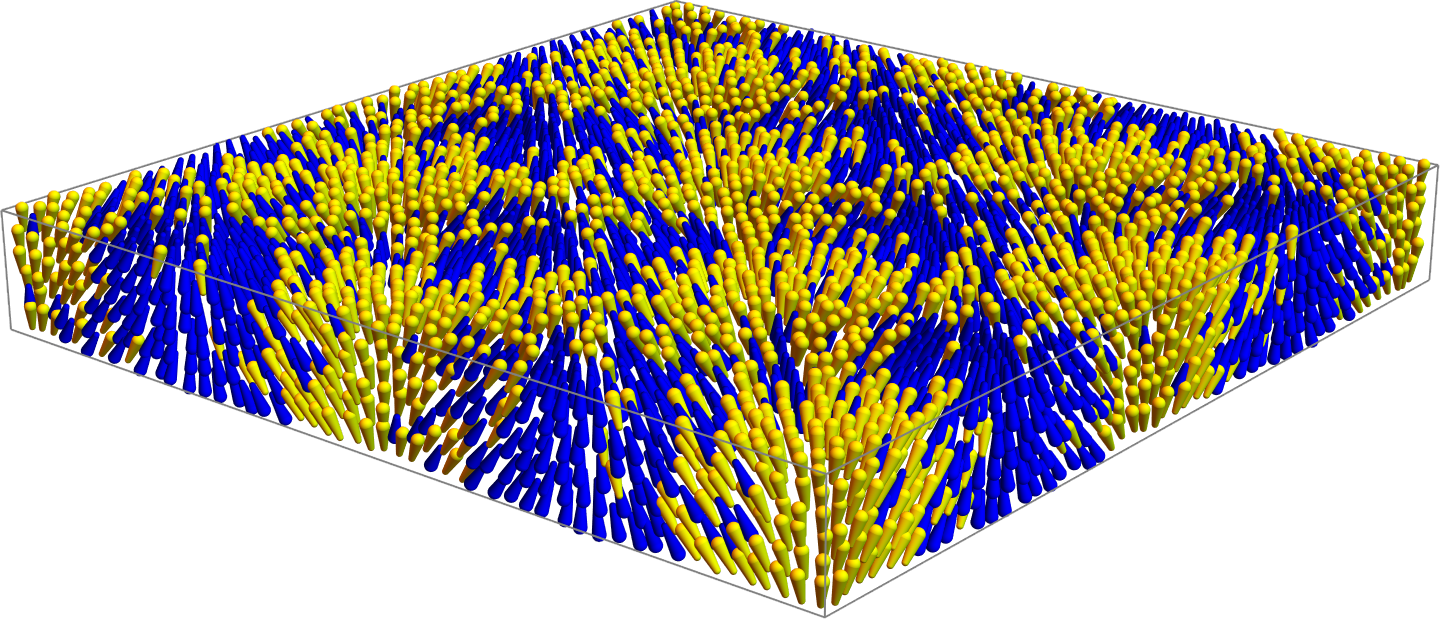}
\caption{Schematic representations of proposed structures for the splay nematic ($N_S$) phase:  (a)~Single splay.  (b)~Double splay.  In both cases, yellow represents regions of positive polar order and positive splay, while blue represents regions of negative polar order and negative splay.}
\end{figure*}

The $N_S$ phase has been modeled\cite{mertelj2018,sebastian2020,copic2020} using the same theoretical method that was previously used for the $N_{TB}$ and $N_{SB}$ phases.  These theoretical studies begin with the understanding that it is impossible to fill space with pure uniform splay, and hence the $N_S$ phase must have a complex structure that includes another deformation mode.  In particular, they assume that the $N_S$ phase has a one-dimensional (1D) modulated structure with alternating regions of splay and bend, as shown in Fig.~1(a).  It has been pointed out that this structure has the same symmetry as the $N_{SB}$ phase.\cite{chaturvedi2019}  We will refer to this structure as \emph{single splay}.

The purpose of this article is to re-examine the structure of the $N_S$ phase in light of a new approach to nematic elasticity theory, which has recently been proposed by Selinger.\cite{selinger2018}  This approach shows that the pure splay deformation is actually \emph{double splay}, in which the director splays inward or outward in two dimensions (2D).  By contrast, single splay is a combination of pure splay with another deformation mode, called biaxial splay or $\Delta$.  The distinction between single and double splay is related to the concept of saddle splay, as discussed in the article.\cite{selinger2018}  If we believe that the $N_S$ phase is induced by a spontaneous splay, then we might expect it to have double splay, rather than single splay.  A double splay structure would have a 2D modulation of the director field, as shown in Fig.~1(b).

To understand the relative stability of single and double splay structures, we begin with the same Landau theory as in previous studies of the $N_S$ phase.\cite{mertelj2018,sebastian2020,copic2020}  We consider possible assumptions for the director field, $\hat{\bf{n}}(x)$ for single splay or $\hat{\bf{n}}(x,y)$ for double splay, insert them into the free energy, average over the volume of the liquid crystal, and minimize over parameters in the assumptions.  We find that the double splay structure generally has a lower free energy than the single splay structure.  It is clearly lower in the critical region near a second-order transition from the nematic to the $N_S$ phase, and it is also lower in most of the low-temperature region.  Single splay is only stable in a small region of the phase diagram at intermediate temperature, where the periodicity of the modulated structure is relatively small.

The plan of this paper is as follows.  In Sec.~\ref{sec:theory}, we present the free energy and analyze the behavior in the critical region near the transition.  In Sec.~\ref{sec:numeric}, we extend the analysis into the lower-temperature region, where the calculations must be done numerically.  This calculation leads to a phase diagram showing uniform nematic, double splay, single splay, and uniform polar phases.  Finally, in Sec.~\ref{sec:discussion}, we discuss the implications of these results for experiments on the $N_S$ phase.

\section{Theory}
\label{sec:theory}

We consider a nematic liquid crystal with director field $\hat{\bf{n}}(\bf{r})$.  If the liquid crystal has polar order $\bf{P}(\bf{r})$ along the director, then the free energy density can be written as
\begin{align}
\label{Eq1}
F = &\frac{1}{2}K_{11} (\nabla \cdot\hat{\bf n})^2 + \frac{1}{2} K_{22} [\hat{\bf n} \cdot (\nabla \times \hat{\bf n})]^2
+ \frac{1}{2} K_{33} |\hat{\bf n} \times (\nabla \times \hat{\bf n})|^2 \nonumber\\
&- \lambda (\nabla \cdot\hat{\bf n})(\hat{\bf n}\cdot{\bf P})
+ \frac{1}{2}\mu |{\bf P}|^2 + \frac{1}{4} \nu |{\bf P}|^4 + \frac{1}{2} \kappa |\nabla {\bf P}|^2.
\end{align}
Here, the first three terms are the Oseen-Frank elastic free energy for director deformations, expressed in terms of the splay, twist, and bend modes.  The fourth term is the flexoelectric coupling between splay and polar order parallel to the director.  The fifth and sixth terms are a standard Landau expansion for the free energy in terms of the polar order parameter.  In this expansion, the quadratic coefficient is assumed to vary linearly with temperature as $\mu(T)=\mu'(T-T_0)$.  The final term gives the free energy penalty for gradients in the polar order.  Note that the free energy is zero in the uniform nematic phase, in which $\hat{\bf n}$ is constant and ${\bf P}=0$.

The free energy of Eq.~\eqref{Eq1} is the same as the free energy used by Mertelj \emph{et al.}\cite{mertelj2018,sebastian2020,copic2020} except for three minor modifications.  First, we express the free energy in terms of the director field $\hat{\bf{n}}(\bf{r})$, while the previous articles use the nematic order tensor $Q_{ij}=S(n_i n_j - \frac{1}{3}\delta_{ij})$.  Because their scalar order parameter $S$ is constant, this is just a change of notation, with no further significance.  Second, we use different letters for some coefficients ($\lambda$ instead of $\gamma$, $\mu$ instead of $t$, $\kappa$ instead of $b$) for consistency with earlier work by our group.\cite{jakli2018,shamid2013}  Once again, this is just a change of notation.  Third, we include the fourth-order term $\frac{1}{4} \nu |{\bf P}|^4$, which the previous articles omit, so that we can assess whether this term is important for the physics.

From the previous work on the $N_S$ phase,\cite{mertelj2018,sebastian2020,copic2020} as well as earlier analogous work on the $N_{TB}$ and $N_{SB}$ phases,\cite{jakli2018,meyer1976,shamid2013} we already know several features of this model.  At high temperatures, the lowest-free-energy state is the uniform nematic phase.  This uniform nematic phase is stable down to a critical point, at which it becomes unstable to coupled fluctuations with nonzero splay $\nabla\cdot\hat{\bf n}$ and nonzero ${\bf P}$.  This critical point occurs when the quadratic coefficient is $\mu_c=\lambda^2 / K_{11}$, corresponding to a critical temperature of $T_c = T_0 + \lambda^2 / (K_{11}\mu')$.  Below the critical point, the system goes into a modulated phase, with a periodic modulation of ${\bf n}$ and nonzero ${\bf P}$.  In the critical regime, where $\delta\mu = \mu_c - \mu$ is small, the amplitude and wavevector of the director modulation both scale as $\delta\mu^{1/2}$, while the amplitude of the polarization modulation scales as $\delta\mu^1$.

Here, we want to consider the structure of the modulated phase below the critical point.  In particular, we consider the following two possible structures.

\subsection{Single splay}

As a first step, we would like to use this free energy to model the single splay state, shown in Fig.~1(a).  In this structure, there is a 1D alternation of two types of domains.  In the yellow domains, the polar order parameter $\bf{P}(\bf{r})$ and the splay vector $\hat{\bf n}(\nabla \cdot\hat{\bf n})$ both point upward, while in the blue domains, these vectors both point downward.  Along the interfaces where opposite domains meet, the polar order goes to zero, as can be seen by the equal populations of yellow and blue.  Also, the splay vanishes along the interfaces, and the local director deformation becomes pure bend.

Near the critical point where the modulation begins, we expect to see smooth sinusoidal variations of the director field and the polar order.  Hence, we assume the director field $\hat{\bf n}(x) = (\sin\theta(x), 0, \cos\theta(x))$, where $\theta(x) = \theta_0 \sin(kx)$.  Because the polar order is coupled to splay, we assume it is aligned with the local director field, so that ${\bf P}(x)=P(x)\hat{\bf n}(x)$, and its magnitude is proportional to the local splay, so that $P(x)=p_0 \nabla\cdot\hat{\bf n}/(k\theta_0)\approx p0\cos(kx)$.  We insert these assumptions for the director field and the polar order parameter into the free energy density~\eqref{Eq1}, and average over the period $2\pi/k$.  Based on the previous work mentioned above, we assume $\theta_0$ and $k$ are both of order $(\mu_c -\mu)^{1/2}$, and $p_0$ is of order $(\mu_c -\mu)^1$, as we will confirm self-consistently.  Hence, we expand all terms to order $(\mu_c -\mu)^4$, and obtain the average free energy
\begin{align}
F_s =& \frac{1}{4}\left[K_{11} k^2 \theta_0^2 - 2\lambda k \theta_0 p_0 + \mu p_0^2 \right]
\left[1-\frac{1}{4}\theta_0^2+\frac{1}{24}\theta_0^4\right]\\
&+\frac{1}{16}K_{33} k^2 \theta_0^4 \left[1-\frac{1}{6}\theta_0^2\right]
+\frac{3}{32}\nu p_0^4
+\frac{1}{4}\kappa k^2 p_0^2 \left[1+\frac{1}{2}\theta_0^2 \right],\nonumber
\end{align}
with the subscript $s$ for single splay.  We then minimize over the variational parameters $\theta_0$, $p_0$, and $k$, and obtain the critical behavior
\begin{align}
& \theta_0 = \frac{2 K_{11}\delta\mu^{1/2}}{(3 K_{33})^{1/2}\lambda}
-\frac{K_{11}^2 (29 K_{11}\kappa\lambda^2 -24 K_{33}\kappa\lambda^2 + 9 K_{11}^3 \nu)\delta\mu^{3/2}}{9 (3 K_{33})^{3/2}\kappa\lambda^5},
\nonumber\\
& p_0 = \frac{2 K_{11}^2 \delta\mu}{3(K_{33}\kappa)^{1/2}\lambda^2}
-\frac{K_{11}^3 (43 K_{11}\kappa\lambda^2 -48 K_{33}\kappa\lambda^2 + 18 K_{11}^3 \nu)\delta\mu^2}{81 (K_{33}\kappa)^{3/2}\lambda^6},
\nonumber\\
& k = \frac{\delta\mu^{1/2}}{(3\kappa)^{1/2}}
-\frac{K_{11} (14 K_{11}\kappa\lambda^2 +12 K_{33}\kappa\lambda^2 + 9 K_{11}^3 \nu)\delta\mu^{3/2}}{18 K_{33}(3\kappa)^{3/2}\lambda^4}.
\label{sspredictions}
\end{align}
Putting these expressions back into the free energy gives the critical behavior
\begin{equation}
F_s = -\frac{K_{11}^4 \delta\mu^3}{27 K_{33}\kappa\lambda^4}
+\frac{K_{11}^5 (26 K_{11}\kappa\lambda^2 - 24 K_{33}\kappa\lambda^2 + 9 K_{11}^3 \nu) \delta\mu^4}{486 K_{33}^2 \kappa^2 \lambda^8}.
\label{Fsingle}
\end{equation}
These results are consistent with previous work on the $N_S$ phase,\cite{mertelj2018,sebastian2020,copic2020} except that we have expanded all the power series to higher order in $\delta\mu = \mu_c - \mu$.  From these results, we can see that the single splay structure has a negative free energy, i.e. lower than the uniform nematic phase, whenever the quadratic coefficient is $\mu<\mu_c$ and hence the temperature is $T<T_c$.  This free energy can be compared with the free energy of an alternative structure.

\subsection{Double splay}

We now consider the double splay state, shown in Fig.~1(b).  In this structure, there is a 2D checkerboard alternation of two types of domains.  Once again, the yellow domains are regions where the polar order parameter $\bf{P}(\bf{r})$ and the splay vector $\hat{\bf n}(\nabla \cdot\hat{\bf n})$ both point upward, and the blue domains are regions where these vectors both point downward.  Along the interfaces where opposite domains meet, the polar order and the splay both go to zero.

The double splay structure can be described mathematically by the director field
\begin{equation}
\hat{\bf n}(x,y) = \frac{(\theta_0\sin(k x)\cos(k y),\theta_0\cos(k x)\sin(k y), 1)}{\sqrt{1 + \theta_0^2 \sin^2(k x) \cos^2(k y) + \theta_0^2 \cos^2(k x)\sin^2(k y)}}.
\end{equation}
As in the previous case, we assume that the polar order is aligned with the director field, ${\bf P}(x,y)=P(x,y)\hat{\bf n}(x,y)$, and its magnitude is proportional to the splay, $P(x,y)=p_0 \nabla\cdot\hat{\bf n}/(k\theta_0) \approx 2 p_0 \cos(k x) \cos(k y)$.  We put these assumptions into the free energy density~\eqref{Eq1}, and average over the periodicity in both $x$ and $y$.  Again,  we assume $\theta_0$ and $k$ are both of order $(\mu_c -\mu)^{1/2}$, while $p_0$ is of order $(\mu_c -\mu)^1$, and expand all terms to order $(\mu_c - \mu)^4$.  This expansion gives the average free energy
\begin{align}
F_d =& \frac{1}{2}\left[K_{11} k^2 \theta_0^2 - 2\lambda k \theta_0 p_0 + \mu p_0^2 \right]
\left[1-\frac{5}{8}\theta_0^2+\frac{15}{32}\theta_0^4\right]\nonumber\\
&+\frac{1}{16}K_{33} k^2 \theta_0^4 \left[1-\frac{5}{4}\theta_0^2\right]
+\frac{9}{16}\nu p_0^4
+\kappa k^2 p_0^2 ,
\end{align}
with the subscript $d$ for double splay.  We minimize over the variational parameters $\theta_0$, $p_0$, and $k$, and obtain the critical behavior
\begin{align}
& \theta_0 = \frac{2^{3/2} K_{11}\delta\mu^{1/2}}{(3 K_{33})^{1/2}\lambda}
+\frac{2^{1/2} K_{11}^2 (5 K_{11}\kappa\lambda^2 +8 K_{33}\kappa\lambda^2 - 9 K_{11}^3 \nu)\delta\mu^{3/2}}{3 (3 K_{33})^{3/2}\kappa\lambda^5},
\nonumber\\
& p_0 = \frac{2 K_{11}^2 \delta\mu}{3(K_{33}\kappa)^{1/2}\lambda^2}
-\frac{K_{11}^3 (5 K_{11}\kappa\lambda^2 -16 K_{33}\kappa\lambda^2 + 18 K_{11}^3 \nu)\delta\mu^2}{27 (K_{33}\kappa)^{3/2}\lambda^6},
\nonumber\\
& k = \frac{\delta\mu^{1/2}}{(6\kappa)^{1/2}}
-\frac{K_{11} (10 K_{11}\kappa\lambda^2 +4 K_{33}\kappa\lambda^2 + 9 K_{11}^3 \nu)\delta\mu^{3/2}}{3 K_{33}(6\kappa)^{3/2}\lambda^4},
\end{align}
with the free energy
\begin{equation}
F_d = -\frac{2 K_{11}^4 \delta\mu^3}{27 K_{33}\kappa\lambda^4}
+\frac{K_{11}^5 (10 K_{11}\kappa\lambda^2 - 8 K_{33}\kappa\lambda^2 + 9 K_{11}^3 \nu) \delta\mu^4}{81 K_{33}^2 \kappa^2 \lambda^8}.
\label{Fdouble}
\end{equation}
These results are similar to the corresponding results for the single splay state, but with different numerical factors.

\subsection{Comparison}

To compare Eqs.~\eqref{Fsingle} and \eqref{Fdouble} for the free energies of the single and double splay structures, we first examine the leading terms of order $\delta\mu^3$.  These terms show that the free energy of the double splay structure is twice as negative as the free energy of the single splay structure (with $F=0$ representing the free energy of the uniform nematic phase).  Hence, the double splay structure is more stable than the single splay structure in the critical regime where $\delta\mu$ is small---i.e.\ the regime where the quadratic coefficient $\mu$ is close to $\mu_c=\lambda^2 / K_{11}$, the temperature $T$ is close to $T_c = T_0 + \lambda^2 / (K_{11}\mu')$, the amplitudes $\theta_0$ and $p_0$ are small, and the wave vector $k$ is small.  This theoretical result agrees with our general expectation that pure splay should be double splay, not single splay, based on the recent approach to nematic elasticity theory.\cite{selinger2018}  In the critical regime, the free energy favors pure splay, and it is not greatly influenced by details of the structure, like the director deformations in the interfaces between positive and negative splay.  For that reason, the free energy prefers the double splay structure compared with the single splay structure.

Away from the critical regime, as $\delta\mu$ becomes larger, the comparison becomes more complicated.  Here, the terms of order $\delta\mu^4$ become important.  In the single and double splay structures, these terms depend on the Frank constants $K_{11}$ and $K_{33}$ in different ways.  They also depend in different ways on the quartic coefficient $\nu$ in the Landau expansion for the free energy of polar order, which is why we need to include this coefficient in the theory.  Based on the $\delta\mu^4$ terms in Eqs.~\eqref{Fsingle} and \eqref{Fdouble}, the single splay structure is favored by a large ratio of Frank constants $K_{11}/K_{33}$ and a large coefficient $\nu$.  Hence, it is possible that the single splay structure might become more stable than the double splay structure farther below the critical point, i.e.\ deeper in the $N_S$ phase, depending on $K_{11}/K_{33}$ and $\nu$.

We would like to assess whether the single splay structure ever becomes more stable than the double splay structure in the regime away from the critical point.  However, comparing the power series expansions is not the best way to make this assessment, because these expansions were derived with the assumption that the director deformation has a simple sinusoidal form.  Away from the critical point, this assumption might not be correct.  Instead, the shape of the deformation might be more complex.  For that reason, we should perform numerical calculations to minimize the free energy for 1D or 2D modulations, and compare the results of the numerical calculations.  These calculations will be presented in the following section.

\section{Numerical solution}
\label{sec:numeric}

For the purpose of numerical calculations, we would like to reduce the number of parameters in the theory.  Hence, without loss of generality, we choose units of energy, length, and polarization to set $\lambda=1$, $K_{11} =1$, and $\kappa=1$.  With this rescaling, the free energy of Eq.~\eqref{Eq1} becomes
\begin{align}
\tilde{F} = &\frac{1}{2} (\nabla \cdot\hat{\bf n})^2 + \frac{1}{2} \tilde{K}_{22} [\hat{\bf n} \cdot (\nabla \times \hat{\bf n})]^2
+ \frac{1}{2} \tilde{K}_{33} |\hat{\bf n} \times (\nabla \times \hat{\bf n})|^2 \nonumber\\
&- (\nabla \cdot\hat{\bf n})(\hat{\bf n}\cdot{\bf P})
+ \frac{1}{2}\tilde{\mu} |{\bf P}|^2 + \frac{1}{4} \tilde{\nu} |{\bf P}|^4 + \frac{1}{2} |\nabla {\bf P}|^2,
\label{Fdimensionless}
\end{align}
in which the dimensionless constants are $\tilde{K}_{22} = K_{22}/K_{11}$, $\tilde{K}_{33} = K_{33}/K_{11}$, $\tilde{\mu} = \mu K_{11}/\lambda^2$, and $\tilde{\nu} = \nu K_{11}^2/(\kappa \lambda^2)$.  Hence, the critical point occurs at $\tilde{\mu}_c = 1$.

\subsection{Single splay}

To model the single splay structure, we consider a director field of the form $\hat{\bf n}(x) = (\sin\theta(x), 0, \cos\theta(x))$ and a polar order parameter of the form ${\bf P}(x)=P(x)\hat{\bf n}(x)$.  Putting those expressions into the scaled free energy gives
\begin{align}
\tilde{F} = &\frac{1}{2}(\cos^2 \theta)\theta'^2
+ \frac{1}{2} \tilde{K}_{33}(\sin^2 \theta)\theta'^2 \\
&- (\cos\theta)\theta' P
+ \frac{1}{2}\tilde{\mu} P^2 + \frac{1}{4} \tilde{\nu} P^4 + \frac{1}{2} (P'^2 + P^2 \theta'^2).\nonumber
\end{align}
Minimizing the free energy over the functions $\theta(x)$ and $P(x)$ then gives the Euler-Lagrange equations
\begin{align}
0 = & (\cos^2 \theta + \tilde{K}_{33} \sin^2 \theta)\theta'' - (1 - \tilde{K}_{33})(\cos\theta\sin\theta)\theta'^2 \nonumber\\
&- (\cos\theta) P' + 2 P P' \theta' + P^2 \theta'', \nonumber\\
0 = & (\cos\theta)\theta' - \tilde{\mu} P - \tilde{\nu} P^3 + P'' - P \theta'^2 ,
\end{align}
respectively.

For any set of energetic parameters $\tilde{\mu}$, $\tilde{\nu}$, and $\tilde{K}_{33}$, we solve the Euler-Lagrange equations numerically.  We expect to find periodic solutions for $\theta(x)$ and $P(x)$, but we do not know the periodicity in advance.  Hence, we solve the Euler-Lagrange equations using periodic boundary conditions on an interval $0\leq x\leq L$, with an arbitrary scaled wavelength $L$.  We put the solutions back into the free energy density~\eqref{Fdimensionless}, integrate over the interval to find the total free energy, and divide by $L$ to find the average free energy density for that $L$.  We then repeat the calculation for different values of $L$, and minimize the average free energy density over $L$.  This procedure gives the optimum $L$ as well as the optimum functions $\theta(x)$ and $P(x)$ and the optimum value of the average free energy density for that set of energetic parameters.

\begin{figure*}
(a)\includegraphics[width=0.47\linewidth]{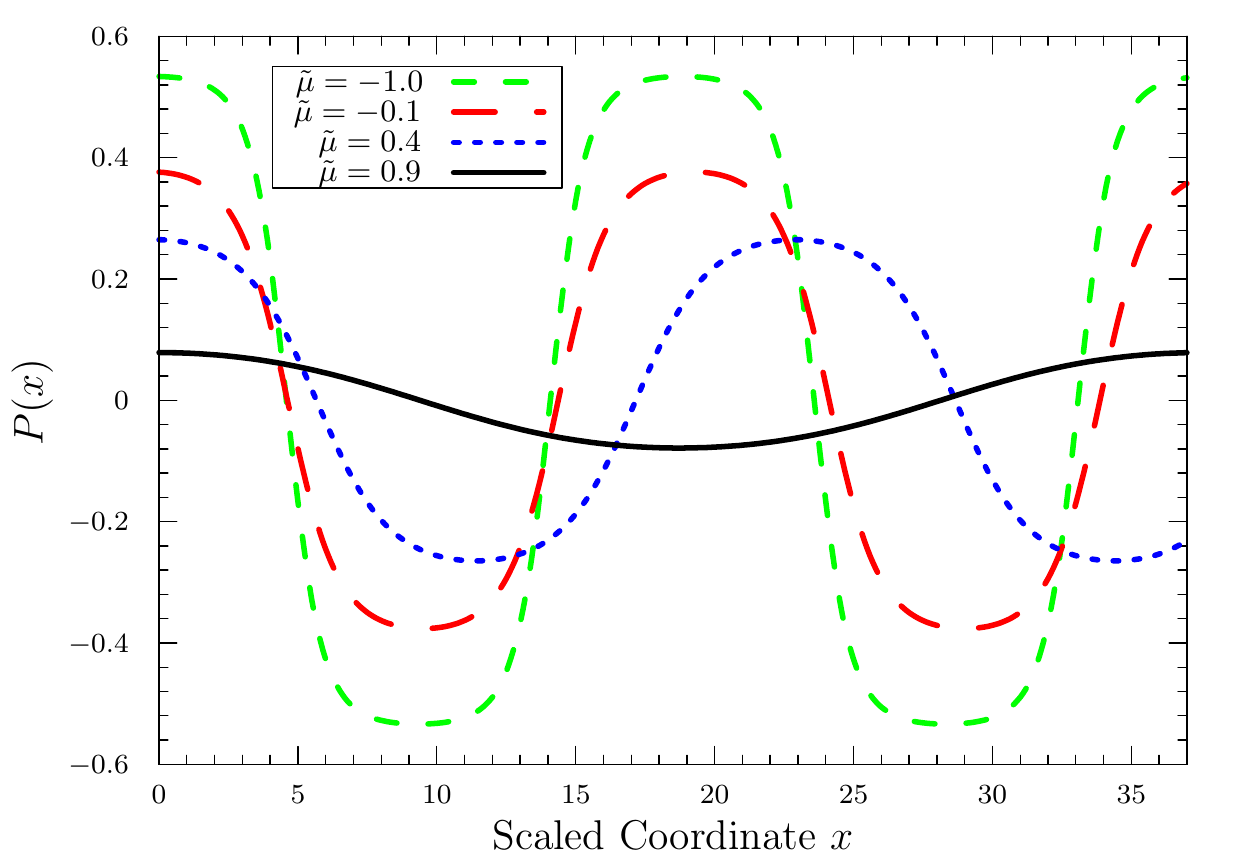}
(b)\includegraphics[width=0.47\linewidth]{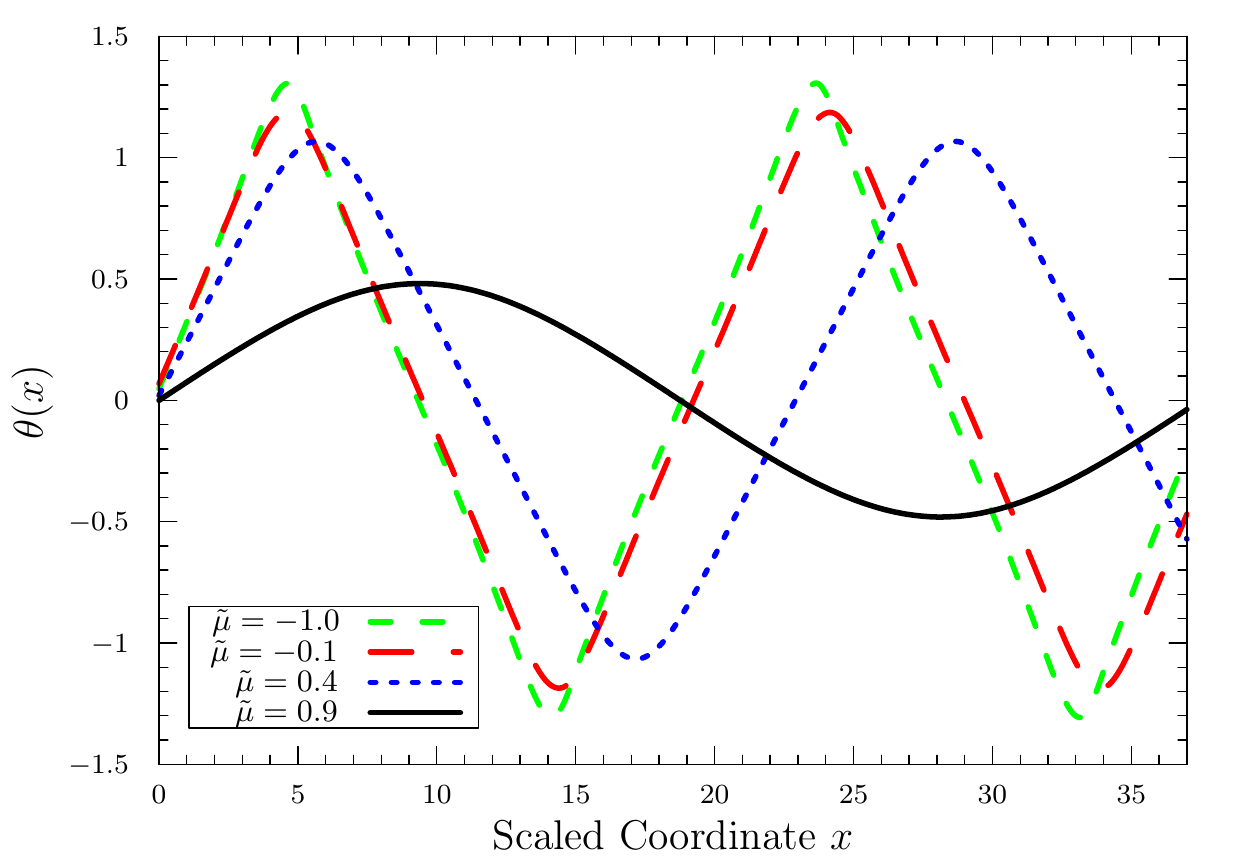}
\caption{Numerical results for (a) the polar order parameter $P(x)$ and (b) the director orientation $\theta(x)$, as functions of the scaled coordinate $x$, for several values of the scaled temperature $\tilde{\mu}$, at fixed parameters $\tilde{\nu} = 5$ and $K_{11}/K_{33}=8$.}
\label{theta-p-graphics}
\end{figure*}

When $\tilde{\mu}$ is slightly below 1 (i.e. temperature $T$ just below the critical temperature $T_c$), the plots of $\theta(x)$ and $P(x)$ are sine and cosine waves, consistent with the assumption in Sec.~2.1.  When $\tilde{\mu}$ decreases further (i.e. $T$ significantly below $T_c$), the amplitudes of these waves grow larger, and the shapes of the waves also change, as shown in Fig.~\ref{theta-p-graphics}.  In that regime, the system forms a series of wide domains separated by relatively narrow walls.  In the positive domains, $P(x)$ is approximately a positive constant, and $\theta(x)$ increases approximately linearly.  In the negative domains, $P(x)$ is approximately a negative constant, and $\theta(x)$ decreases approximately linearly.  Across each wall, $P(x)$ changes sign in a narrow region, and $\theta(x)$ is approximately constant.  These walls can be regarded as solitons in the polar order.  The temperature-dependent crossover from sinusoidal stripes to a soliton-like modulation is similar to the behavior predicted for chiral stripes in Langmuir monolayers and smectic films.\cite{selinger1993} 

\begin{figure}
\includegraphics[width=\columnwidth]{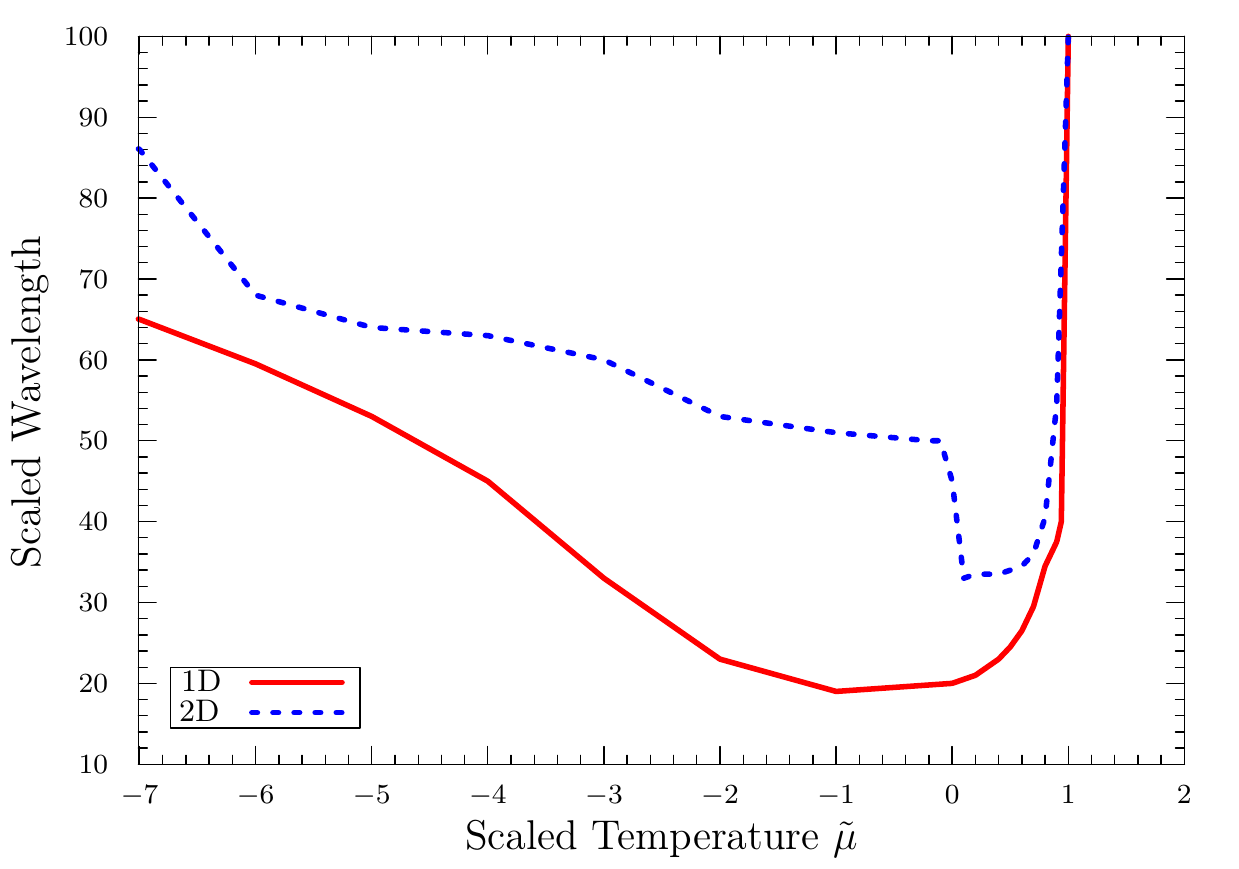}
\caption{Scaled wavelength as a function of scaled temperature $\tilde{\mu}$, for fixed parameters $\tilde{\nu} = 5$ and $K_{11}/K_{33}=8$.  The red line denotes the single splay modulation, while the blue line represents the double splay modulation.}
\label{wavelength}
\end{figure}

We have done a series of numerical calculations with varying $\tilde{\mu}$, corresponding to varying temperature, for fixed ratio $\tilde{K}_{33}^{-1}=K_{11}/K_{33}=8$ and fixed parameter $\tilde{\nu}=5$.  We choose these large values because they should favor the stability of single splay relative to double splay, based on the analysis in Sec.~2.3 of the power series expansions.  In Fig.~\ref{wavelength}, the red line shows the optimal scaled wavelength $L$ as a function of $\tilde{\mu}$.  At the critical point $\tilde{\mu}_c = 1$, the wavelength is infinite.  As $\tilde{\mu}$ decreases into the $N_S$ phase, the wavelength decreases rapidly, as expected from Eq.~\eqref{sspredictions}.  Interestingly, the wavelength reaches a minimum, and then the variation reverses.  When $\tilde{\mu}$ is large and negative, deep in the soliton-like regime, a further decrease in $\tilde{\mu}$ cases the wavelength to increase, so that each domain becomes larger and the walls are farther apart.

\begin{figure}
\includegraphics[width=\columnwidth]{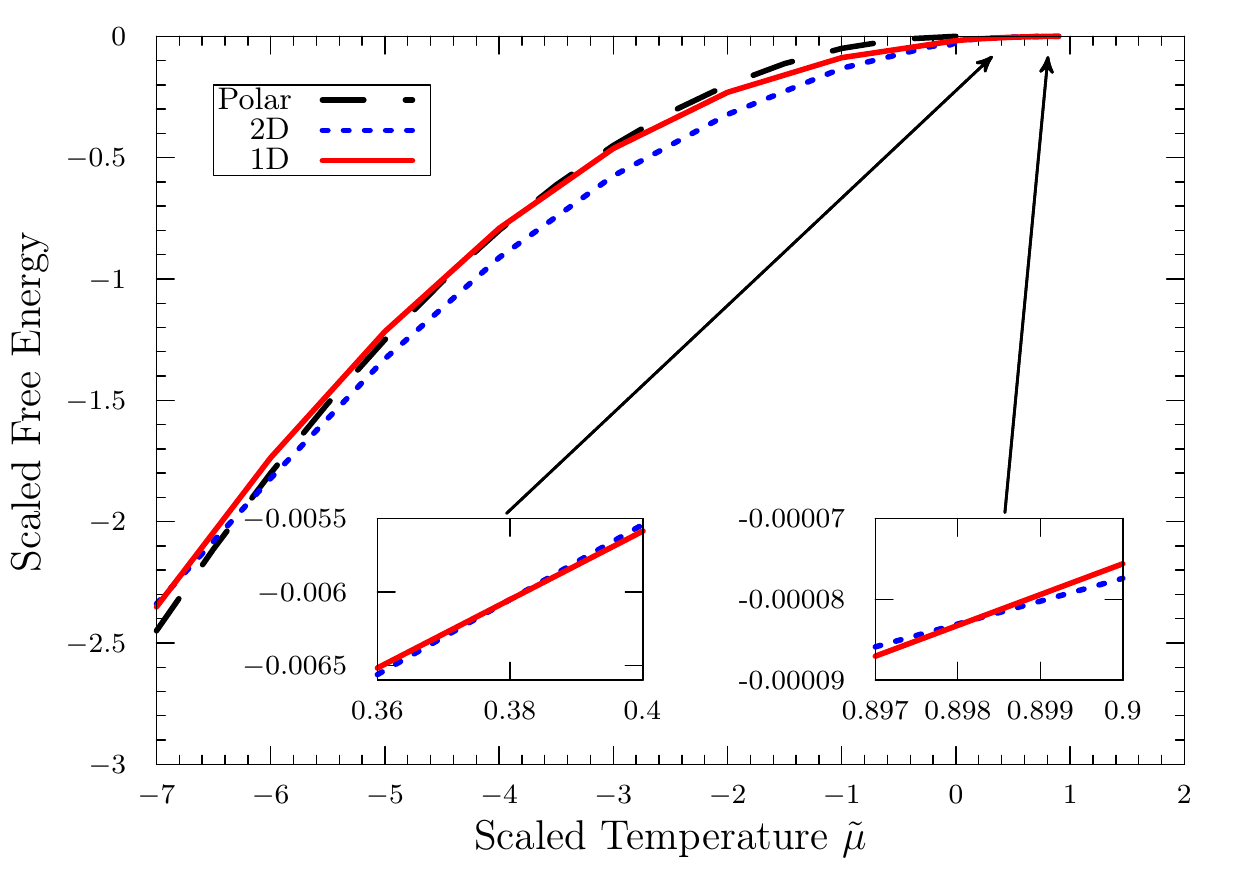}
\caption{Average scaled free energy as a function of scaled temperature $\tilde{\mu}$, for fixed parameters $\tilde{\nu} = 5$ and $K_{11}/K_{33}=8$.  The red line represents the single splay modulation, blue line the double splay modulation, and black line the polar phase.  The insets show the crossover points, with first-order transitions between double splay and single splay.}
\label{free_energy}
\end{figure}

In Fig.~\ref{free_energy}, the red line shows the corresponding result for the average free energy density of the single splay structure, at the optimum wavelength $L$, as a function of $\tilde{\mu}$.  At the critical point, the free energy is zero, which is the free energy of the uniform nematic phase.  As $\tilde{\mu}$ decreases, the free energy becomes more negative.  This free energy result can be compared with alternative structures.

\subsection{Double splay}

To model the double splay structure, we consider the director field
\begin{equation}
\hat{\bf n}(x,y)=\left(n_x(x,y),n_y(x,y),[1-n_x(x,y)^2-n_y(x,y)^2]^{1/2}\right),
\end{equation}
along with the polar order parameter ${\bf P}(x,y)=P(x,y)\hat{\bf n}(x,y)$.  We put those expressions into the Eq.~\eqref{Fdimensionless}, to obtain the scaled free energy density in terms of the three functions $n_x(x,y)$, $n_y(x,y)$, and $P(x,y)$.  We then minimize the free energy over those three functions, and obtain three coupled 2D Euler-Lagrange equations.  The equations are too lengthy to reproduce here.

For any set of energetic parameters $\tilde{\mu}$, $\tilde{\nu}$, and $\tilde{K}_{33}$, we solve the 2D Euler-Lagrange equations numerically on the square domain $0\leq x\leq L$, $0\leq y\leq L$, using periodic boundary conditions.  We put the solutions back into the free energy density, integrate over the square domain, and divide by $L^2$ to find the average free energy density for that $L$.  We then repeat the calculation to minimize the average free energy density over $L$.  In this way, we obtain the optimum $L$, the optimum functions $n_x(x,y)$, $n_y(x,y)$, and $P(x,y)$, and the optimum free energy for that set of energetic parameters.

\begin{figure*}
(a)\includegraphics[width=0.47\linewidth]{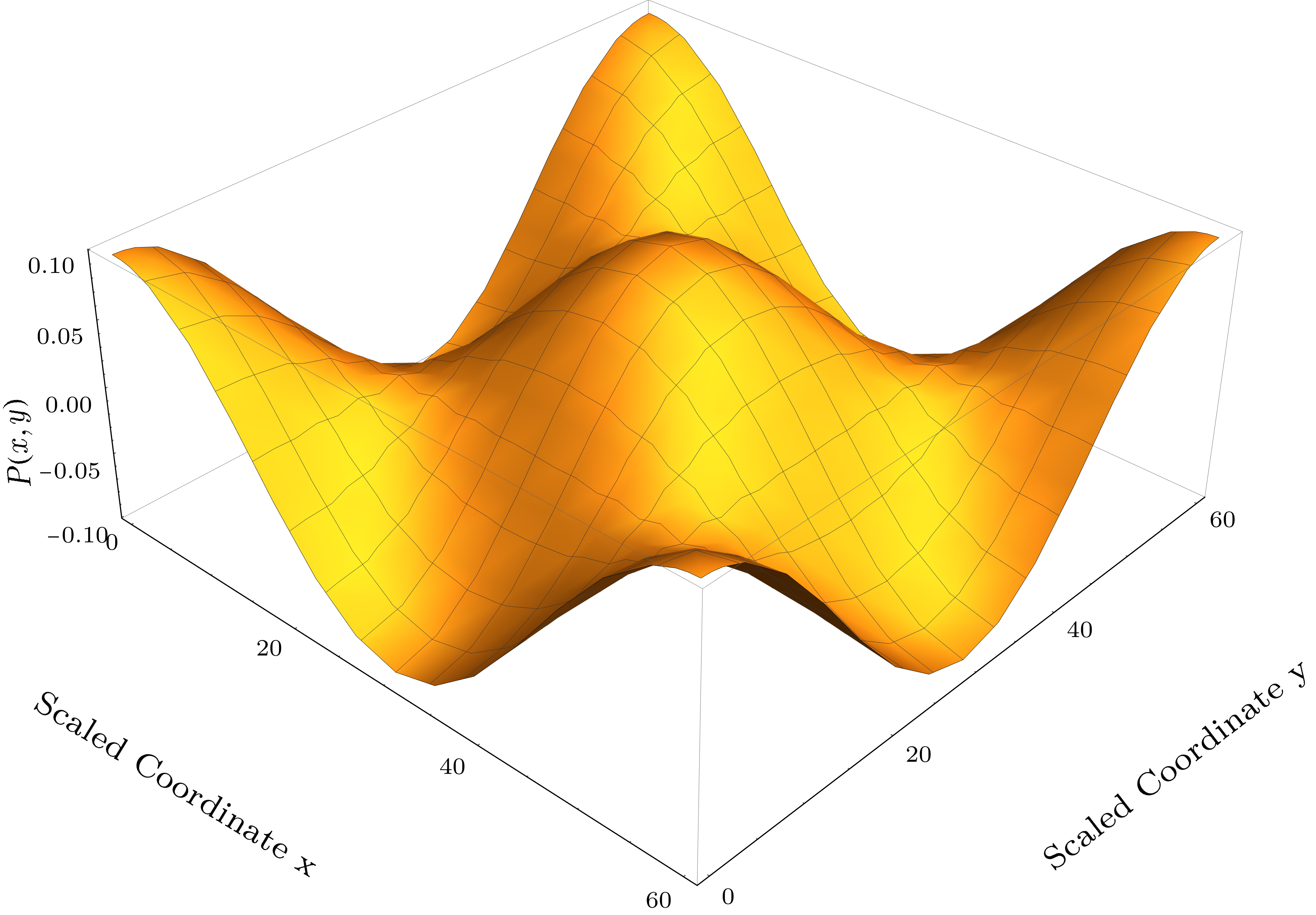}
(b)\includegraphics[width=0.47\linewidth]{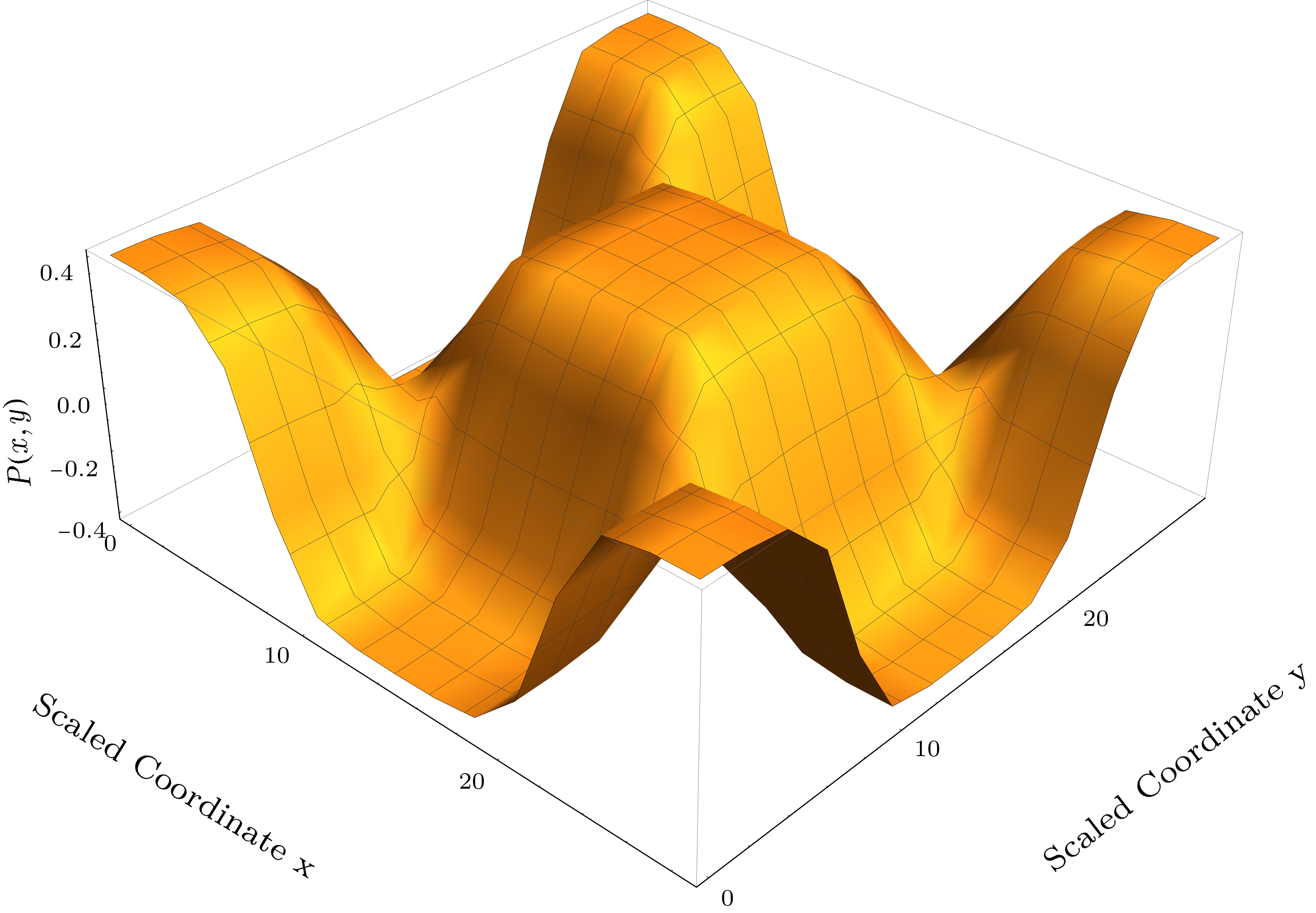}
\caption{Numerical results for the polar order parameter $P(x,y)$ for two values of the scaled temperature, (a)~$\tilde{\mu}=0.9$ and (b)~$\tilde{\mu}=-0.1$. In both cases, $\tilde{\nu} = 5$ and $K_{11}/K_{33}=8$.}
\label{P(x,y)}
\end{figure*}

The numerical results show that the shape of $n_x(x,y)$, $n_y(x,y)$, and $P(x,y)$ changes as a function of $\tilde{\mu}$, in the same way as in the single splay case.  Two examples are presented in Fig.~\ref{P(x,y)}.  When $\tilde{\mu}$ is slightly below 1 ($T$ just below $T_c$), these plots are sine and cosine waves, as assumed in Sec.~2.2.  When $\tilde{\mu}$ decreases further ($T$ just below $T_c$), the system forms a lattice of well-defined square domains separated by sharp, soliton-like walls.  In each square domain, $P(x,y)$ is approximately a positive or negative constant, and the director field shows outward or inward double splay.  Across each wall, $P(x,y)$ changes sign and the director is approximately constant.

The wavelength also changes as a function of $\tilde{\mu}$, in approximately the same way as in the single splay case.  In Fig.~\ref{wavelength}, the blue line shows the scaled wavelength of the double splay structure, for fixed parameters $\tilde{\nu} = 5$ and $K_{11}/K_{33}=8$.  This wavelength is infinite at the critical point $\tilde{\mu}_c = 1$, and it decreases rapidly as $\tilde{\mu}$ decreases into the $N_S$ phase.  Eventually it reaches a minimum, and then the variation reverses.  In the soliton-like regime, for large negative $\tilde{\mu}$, the wavelength increases as a function of decreasing $\tilde{\mu}$ (decreasing temperature).  The numerical results are not as smooth as in the single splay case, presumably because the numerical algorithm has more difficulty with the 2D than 1D Euler-Lagrange equations.

We also calculate the average free energy density of the double splay structure as a function of $\tilde{\mu}$.  These results are shown by the blue line in Fig.~\ref{free_energy}, for comparison with other structures.

\subsection{Uniform polar phase}

In addition to the single splay and double splay states, we should also consider the possibility of a uniform polar phase.  In this phase, the director $\hat{\bf n}$ is uniform, and the polar order parameter ${\bf P}=P\hat{\bf n}$ is uniform and  nonzero.  Because all gradients vanish, the scaled free energy of Eq.~\eqref{Fdimensionless} becomes just $\tilde{F} = \frac{1}{2}\tilde{\mu}P^2 + \frac{1}{4} \tilde{\nu}P^4$.  Minimizing over $P$ gives $P=0$ for $\tilde{\mu}>0$, and $P=(-\tilde{\mu}/\tilde{\nu})^{1/2}$ for $\tilde{\mu}<0$.  Putting that solution back into the free energy then gives
\begin{equation}
\tilde{F}=
\begin{cases}
0, &\mbox{for } \tilde{\mu}>0, \\ 
-\tilde{\mu}^2 /(4\tilde{\nu}), & \mbox{for } \tilde{\mu}<0.
\end{cases}
\end{equation}
This free energy is shown by the black line in Fig.~\ref{free_energy}.  We did not need to consider this phase in Sec.~2, when we were concentrating on the behavior close to the critical point, for $\tilde{\mu}$ slightly below 1.  However, we must consider it now, because we are studying a wider range of $\tilde{\mu}$, corresponding to lower temperatures deep in the $N_S$ phase.

\subsection{Comparison}

We can now compare the free energies of the single splay, double splay, and uniform polar states.  Figure~\ref{free_energy} shows all three free energies as functions of $\tilde{\mu}$, for fixed $\tilde{\nu} = 5$ and $K_{11}/K_{33}=8$.  The free energies are very close together, but we can see some transitions as $\tilde{\mu}$ is reduced (i.e.\ as the temperature decreases).  Just below the critical point, for $\tilde{\mu}$ slightly less than 1, the double splay structure has the lowest free energy.  This result is consistent with the results of the power series expansions in Sec.~2.  At $\tilde{\mu}\approx 0.898$, the free energy of the single splay structure becomes the lowest, and hence the system has a first-order transition from double splay to single splay.  At $\tilde{\mu}\approx 0.38$, the double splay free energy becomes the lowest once again, and the system has a first-order transition from single splay back to double splay.  (Because those two crossings of the free energy curves are difficult to see in the main plot, they are highlighted in insets.)  The double splay state remains the stable phase over a wide range of $\tilde{\mu}$.  Finally, at $\tilde{\mu}\approx -6.2$, the system has a transition from double splay into the uniform polar phase.

\begin{figure}
\centering
\includegraphics[width=\columnwidth]{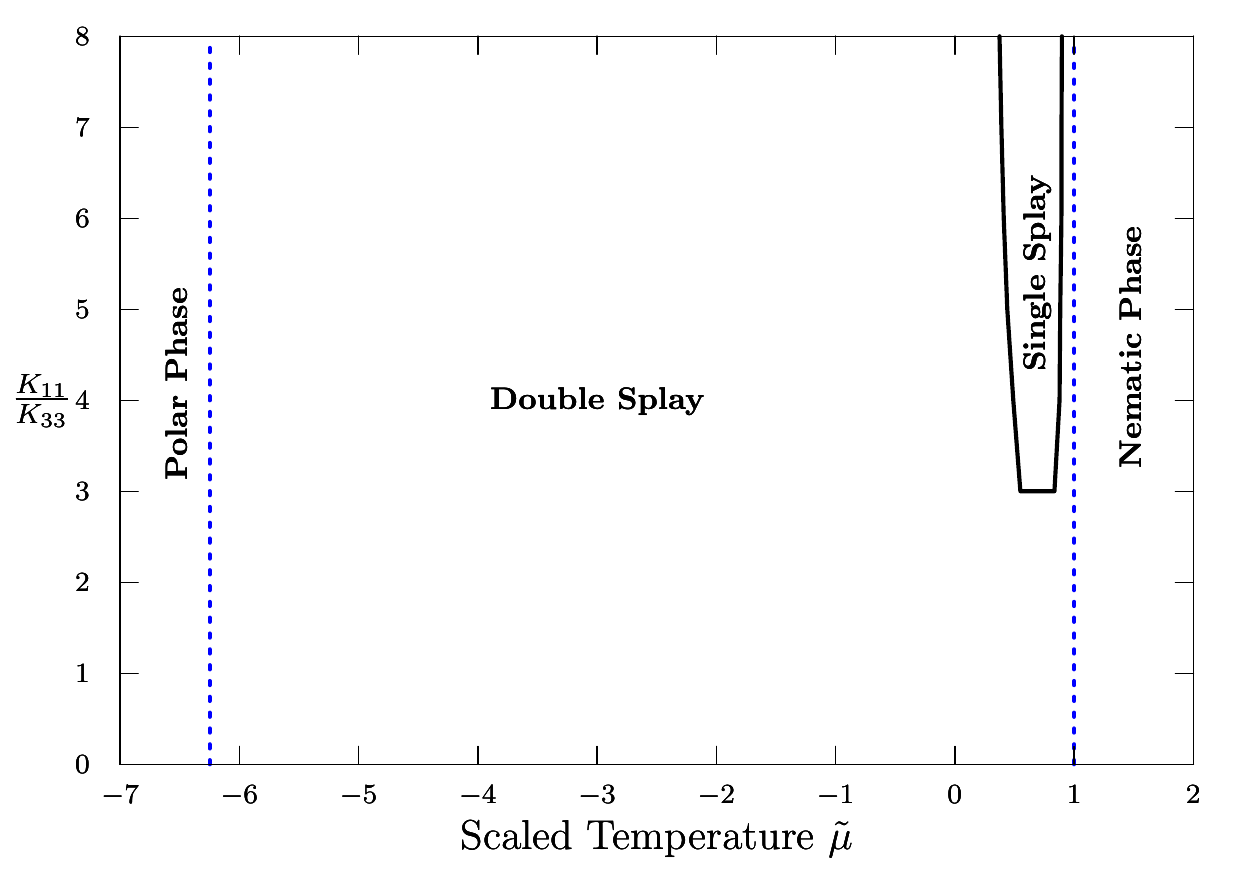}
\caption{Phase diagram for the model studied in this article, as a function of the scaled temperature $\tilde{\mu}$ and the ratio $K_{11}/K_{33}$, for fixed parameter $\tilde{\nu} = 5$.}
\label{diagram}
\end{figure}

We have repeated the calculation for different values of $K_{11}/K_{33}$, and the results are summarized in the phase diagram of Fig.~\ref{diagram}.  For large ratio $K_{11}/K_{33}$, the phase diagram shows the same series of phases described above:  the uniform nematic phase at high $\tilde{\mu}$, a transition to double splay at the critical point $\tilde{\mu}_c=1$, a modest window of single splay in a narrow range of $\tilde{\mu}$, and finally a low-temperature transition into the uniform polar phase.  When $K_{11}/K_{33}$ decreases, the window of single splay becomes narrower and then disappears, so that the only phases are uniform nematic, double splay, and uniform polar.  This result agrees with our expectation in Sec.~2.3, based on the power series expansion, that high $K_{11}/K_{33}$ would help to stabilize single splay compared with double splay.  We have not yet determined how the phase diagram depends on the parameter $\tilde{\nu}$.

One might ask why the single splay state occurs in a narrow window of $\tilde{\mu}$.  To address that question, we separately calculate all the different terms of the free energy, using the numerical solutions for single splay and double splay states.  We find numerically that the term $ \frac{1}{2} \kappa |\nabla {\bf P}|^2$ favors single splay compared with double splay.  This term involves the first derivative of polar order, and polar order is coupled with splay of the director field, and hence this term is essentially a second derivative of the director field.  Because it is a second derivative, it is large whenever the wavelength of the director modulation is small.  We note that the window of stability for single splay is approximately the same as the range of $\tilde{\mu}$ in which the wavelength is smallest.  Hence, we speculate that the small wavelength is responsible for the single splay state.

\section{Discussion}
\label{sec:discussion}

In this paper, we investigate the structure of the $N_S$ phase that is predicted by the free energy of Eq.~\eqref{Eq1}.  This is the simplest free energy that couples splay and polar order, and it is the same free energy that was used in previous studies of the $N_S$ phase.\cite{mertelj2018,sebastian2020,copic2020}  We find that the ground state of this free energy is normally the double splay rather than the single splay structure.  Power series calculations show that double splay has a lower free energy than single splay in the critical regime, and numerical calculations show that double splay has the lowest free energy in most of the phase diagram.  Single splay is only the ground state in a narrow range of temperature, where the predicted wavelength is relatively short.

A further challenge is how to reconcile this theoretical prediction with experimental studies of the $N_S$ phase, which report only a 1D director modulation.  We can suggest three possibilities.

First, perhaps the experiments really do have a 2D double splay modulation, which has not been noticed yet.  This might occur, for example, if the modulation wavelength is larger than or comparable to the thickness of the experimental cell.  In that case, boundary conditions on the top and bottom surfaces might suppress the modulation in one direction, so that only the perpendicular modulation can be observed.  One priority for experiments should be to examine this possibility.

Second, perhaps the experiments have always been done in the region of the phase diagram that has a 1D single splay modulation.  This possibility cannot be ruled out.  However, it seems rather unlikely, considering that the single splay region of the phase diagram is fairly small.

Third, the most interesting theoretical possibility is that some important physics has not yet been included in the free energy of Eq.~\eqref{Eq1}.  For example, perhaps the optimal packing of molecules has single splay rather than double splay.  This could occur if the molecules have a tendency toward biaxial nematic order, and this biaxial order favors splay in a certain plane.  In that case, the Landau theory should be generalized to include the biaxial order parameter and its coupling with director modulations.  This generalization is a promising route for future theoretical research.

\section*{Conflicts of interest}

There are no conflicts to declare.

\section*{Acknowledgements}

This work was funded in part by NSF Grant No. DMR-1409658 and by Coordena\c{c}\~ao de Aperfei\c{c}oamento de Pessoal de N\'ivel Superior - Brazil (CAPES) - Finance Code 001.

\bibliography{ref} 

\providecommand*{\mcitethebibliography}{\thebibliography}
\csname @ifundefined\endcsname{endmcitethebibliography}
{\let\endmcitethebibliography\endthebibliography}{}
\begin{mcitethebibliography}{29}
\providecommand*{\natexlab}[1]{#1}
\providecommand*{\mciteSetBstSublistMode}[1]{}
\providecommand*{\mciteSetBstMaxWidthForm}[2]{}
\providecommand*{\mciteBstWouldAddEndPuncttrue}
  {\def\EndOfBibitem{\unskip.}}
\providecommand*{\mciteBstWouldAddEndPunctfalse}
  {\let\EndOfBibitem\relax}
\providecommand*{\mciteSetBstMidEndSepPunct}[3]{}
\providecommand*{\mciteSetBstSublistLabelBeginEnd}[3]{}
\providecommand*{\EndOfBibitem}{}
\mciteSetBstSublistMode{f}
\mciteSetBstMaxWidthForm{subitem}
{(\emph{\alph{mcitesubitemcount}})}
\mciteSetBstSublistLabelBeginEnd{\mcitemaxwidthsubitemform\space}
{\relax}{\relax}

\bibitem[J{\'{a}}kli \emph{et~al.}(2018)J{\'{a}}kli, Lavrentovich, and
  Selinger]{jakli2018}
A.~J{\'{a}}kli, O.~D. Lavrentovich and J.~V. Selinger, \emph{Rev. Mod. Phys.},
  2018, \textbf{90}, 045004\relax
\mciteBstWouldAddEndPuncttrue
\mciteSetBstMidEndSepPunct{\mcitedefaultmidpunct}
{\mcitedefaultendpunct}{\mcitedefaultseppunct}\relax
\EndOfBibitem
\bibitem[Meyer(1969)]{meyer1969}
R.~B. Meyer, \emph{Phys. Rev. Lett.}, 1969, \textbf{22}, 918--921\relax
\mciteBstWouldAddEndPuncttrue
\mciteSetBstMidEndSepPunct{\mcitedefaultmidpunct}
{\mcitedefaultendpunct}{\mcitedefaultseppunct}\relax
\EndOfBibitem
\bibitem[Prost and Marcerou(1977)]{prost1977}
J.~Prost and J.~Marcerou, \emph{Journal de Physique}, 1977, \textbf{38},
  315--324\relax
\mciteBstWouldAddEndPuncttrue
\mciteSetBstMidEndSepPunct{\mcitedefaultmidpunct}
{\mcitedefaultendpunct}{\mcitedefaultseppunct}\relax
\EndOfBibitem
\bibitem[Harden \emph{et~al.}(2006)Harden, Mbanga, {\'{E}}ber, Fodor-Csorba,
  Sprunt, Gleeson, and J{\'{a}}kli]{harden2006}
J.~Harden, B.~Mbanga, N.~{\'{E}}ber, K.~Fodor-Csorba, S.~Sprunt, J.~T. Gleeson
  and A.~J{\'{a}}kli, \emph{Phys. Rev. Lett.}, 2006, \textbf{97}, 157802\relax
\mciteBstWouldAddEndPuncttrue
\mciteSetBstMidEndSepPunct{\mcitedefaultmidpunct}
{\mcitedefaultendpunct}{\mcitedefaultseppunct}\relax
\EndOfBibitem
\bibitem[Meyer(1976)]{meyer1976}
R.~B. Meyer, in \emph{Molecular Fluids (Les Houches Summer School in
  Theoretical Physics, 1973)}, ed. R.~Balian and G.~Weill, Gordon and Breach,
  New York, 1976, pp. 271--343\relax
\mciteBstWouldAddEndPuncttrue
\mciteSetBstMidEndSepPunct{\mcitedefaultmidpunct}
{\mcitedefaultendpunct}{\mcitedefaultseppunct}\relax
\EndOfBibitem
\bibitem[Dozov(2001)]{dozov2001}
I.~Dozov, \emph{EPL}, 2001, \textbf{56}, 247--253\relax
\mciteBstWouldAddEndPuncttrue
\mciteSetBstMidEndSepPunct{\mcitedefaultmidpunct}
{\mcitedefaultendpunct}{\mcitedefaultseppunct}\relax
\EndOfBibitem
\bibitem[Memmer(2002)]{memmer2002}
R.~Memmer, \emph{Liq. Cryst.}, 2002, \textbf{29}, 483--496\relax
\mciteBstWouldAddEndPuncttrue
\mciteSetBstMidEndSepPunct{\mcitedefaultmidpunct}
{\mcitedefaultendpunct}{\mcitedefaultseppunct}\relax
\EndOfBibitem
\bibitem[Shamid \emph{et~al.}(2013)Shamid, Dhakal, and Selinger]{shamid2013}
S.~M. Shamid, S.~Dhakal and J.~V. Selinger, \emph{Phys. Rev. E}, 2013,
  \textbf{87}, 052503\relax
\mciteBstWouldAddEndPuncttrue
\mciteSetBstMidEndSepPunct{\mcitedefaultmidpunct}
{\mcitedefaultendpunct}{\mcitedefaultseppunct}\relax
\EndOfBibitem
\bibitem[Meyer \emph{et~al.}(2013)Meyer, Luckhurst, and Dozov]{meyer2013}
C.~Meyer, G.~R. Luckhurst and I.~Dozov, \emph{Physical Review Letters}, 2013,
  \textbf{111}, 067801\relax
\mciteBstWouldAddEndPuncttrue
\mciteSetBstMidEndSepPunct{\mcitedefaultmidpunct}
{\mcitedefaultendpunct}{\mcitedefaultseppunct}\relax
\EndOfBibitem
\bibitem[Virga(2014)]{virga2014}
E.~G. Virga, \emph{Phys. Rev. E}, 2014, \textbf{89}, 052502\relax
\mciteBstWouldAddEndPuncttrue
\mciteSetBstMidEndSepPunct{\mcitedefaultmidpunct}
{\mcitedefaultendpunct}{\mcitedefaultseppunct}\relax
\EndOfBibitem
\bibitem[Barbero \emph{et~al.}(2015)Barbero, Evangelista, Rosseto, Zola, and
  Lelidis]{barbero2015}
G.~Barbero, L.~R. Evangelista, M.~P. Rosseto, R.~S. Zola and I.~Lelidis,
  \emph{Phys. Rev. E}, 2015, \textbf{92}, 030501\relax
\mciteBstWouldAddEndPuncttrue
\mciteSetBstMidEndSepPunct{\mcitedefaultmidpunct}
{\mcitedefaultendpunct}{\mcitedefaultseppunct}\relax
\EndOfBibitem
\bibitem[Vaupoti{\v{c}} \emph{et~al.}(2016)Vaupoti{\v{c}}, Curk, Osipov,
  {\v{C}}epi{\v{c}}, Takezoe, and Gorecka]{vaupotic2016}
N.~Vaupoti{\v{c}}, S.~Curk, M.~A. Osipov, M.~{\v{C}}epi{\v{c}}, H.~Takezoe and
  E.~Gorecka, \emph{Phys. Rev. E}, 2016, \textbf{93}, 022704\relax
\mciteBstWouldAddEndPuncttrue
\mciteSetBstMidEndSepPunct{\mcitedefaultmidpunct}
{\mcitedefaultendpunct}{\mcitedefaultseppunct}\relax
\EndOfBibitem
\bibitem[Meyer and Dozov(2016)]{meyer2016}
C.~Meyer and I.~Dozov, \emph{Soft Matter}, 2016, \textbf{12}, 574--580\relax
\mciteBstWouldAddEndPuncttrue
\mciteSetBstMidEndSepPunct{\mcitedefaultmidpunct}
{\mcitedefaultendpunct}{\mcitedefaultseppunct}\relax
\EndOfBibitem
\bibitem[Chen \emph{et~al.}(2013)Chen, Porada, Hooper, Klittnick, Shen,
  Tuchband, Korblova, Bedrov, Walba, Glaser, Maclennan, and Clark]{chen2013}
D.~Chen, J.~H. Porada, J.~B. Hooper, A.~Klittnick, Y.~Shen, M.~R. Tuchband,
  E.~Korblova, D.~Bedrov, D.~M. Walba, M.~A. Glaser, J.~E. Maclennan and N.~A.
  Clark, \emph{Proc. Natl. Acad. Sci. U.S.A.}, 2013, \textbf{110},
  15931--15936\relax
\mciteBstWouldAddEndPuncttrue
\mciteSetBstMidEndSepPunct{\mcitedefaultmidpunct}
{\mcitedefaultendpunct}{\mcitedefaultseppunct}\relax
\EndOfBibitem
\bibitem[Borshch \emph{et~al.}(2013)Borshch, Kim, Xiang, Gao, J{\'{a}}kli,
  Panov, Vij, Imrie, Tamba, Mehl, and Lavrentovich]{borshch2013}
V.~Borshch, Y.-K. Kim, J.~Xiang, M.~Gao, A.~J{\'{a}}kli, V.~P. Panov, J.~K.
  Vij, C.~T. Imrie, M.~G. Tamba, G.~H. Mehl and O.~D. Lavrentovich, \emph{Nat.
  Commun.}, 2013, \textbf{4}, 2365\relax
\mciteBstWouldAddEndPuncttrue
\mciteSetBstMidEndSepPunct{\mcitedefaultmidpunct}
{\mcitedefaultendpunct}{\mcitedefaultseppunct}\relax
\EndOfBibitem
\bibitem[Adlem \emph{et~al.}(2013)Adlem, {\v{C}}opi{\v{c}}, Luckhurst, Mertelj,
  Parri, Richardson, Snow, Timimi, Tuffin, and Wilkes]{adlem2013}
K.~Adlem, M.~{\v{C}}opi{\v{c}}, G.~R. Luckhurst, A.~Mertelj, O.~Parri, R.~M.
  Richardson, B.~D. Snow, B.~A. Timimi, R.~P. Tuffin and D.~Wilkes, \emph{Phys.
  Rev. E}, 2013, \textbf{88}, 022503\relax
\mciteBstWouldAddEndPuncttrue
\mciteSetBstMidEndSepPunct{\mcitedefaultmidpunct}
{\mcitedefaultendpunct}{\mcitedefaultseppunct}\relax
\EndOfBibitem
\bibitem[Chen \emph{et~al.}(2014)Chen, Nakata, Shao, Tuchband, Shuai,
  Baumeister, Weissflog, Walba, Glaser, Maclennan, and Clark]{chen2014}
D.~Chen, M.~Nakata, R.~Shao, M.~R. Tuchband, M.~Shuai, U.~Baumeister,
  W.~Weissflog, D.~M. Walba, M.~A. Glaser, J.~E. Maclennan and N.~A. Clark,
  \emph{Phys. Rev. E}, 2014, \textbf{89}, 022506\relax
\mciteBstWouldAddEndPuncttrue
\mciteSetBstMidEndSepPunct{\mcitedefaultmidpunct}
{\mcitedefaultendpunct}{\mcitedefaultseppunct}\relax
\EndOfBibitem
\bibitem[Meyer \emph{et~al.}(2015)Meyer, Luckhurst, and Dozov]{meyer2015}
C.~Meyer, G.~R. Luckhurst and I.~Dozov, \emph{J. Mater. Chem. C}, 2015,
  \textbf{3}, 318--328\relax
\mciteBstWouldAddEndPuncttrue
\mciteSetBstMidEndSepPunct{\mcitedefaultmidpunct}
{\mcitedefaultendpunct}{\mcitedefaultseppunct}\relax
\EndOfBibitem
\bibitem[Gorecka \emph{et~al.}(2015)Gorecka, Vaupoti{\v{c}}, Zep, Pociecha,
  Yoshioka, Yamamoto, and Takezoe]{gorecka2015}
E.~Gorecka, N.~Vaupoti{\v{c}}, A.~Zep, D.~Pociecha, J.~Yoshioka, J.~Yamamoto
  and H.~Takezoe, \emph{Angew. Chem. Int. Ed.}, 2015, \textbf{54},
  10155--10159\relax
\mciteBstWouldAddEndPuncttrue
\mciteSetBstMidEndSepPunct{\mcitedefaultmidpunct}
{\mcitedefaultendpunct}{\mcitedefaultseppunct}\relax
\EndOfBibitem
\bibitem[Wang \emph{et~al.}(2016)Wang, Zheng, Bisoyi, Gutierrez-Cuevas, Wang,
  Zola, and Li]{wang2016}
Y.~Wang, Z.-G. Zheng, H.~K. Bisoyi, K.~G. Gutierrez-Cuevas, L.~Wang, R.~S. Zola
  and Q.~Li, \emph{Mater. Horiz.}, 2016, \textbf{3}, 442--446\relax
\mciteBstWouldAddEndPuncttrue
\mciteSetBstMidEndSepPunct{\mcitedefaultmidpunct}
{\mcitedefaultendpunct}{\mcitedefaultseppunct}\relax
\EndOfBibitem
\bibitem[Dhakal and Selinger(2010)]{Dhakal2010}
S.~Dhakal and J.~V. Selinger, \emph{Phys. Rev. E}, 2010, \textbf{81},
  031704\relax
\mciteBstWouldAddEndPuncttrue
\mciteSetBstMidEndSepPunct{\mcitedefaultmidpunct}
{\mcitedefaultendpunct}{\mcitedefaultseppunct}\relax
\EndOfBibitem
\bibitem[Mertelj \emph{et~al.}(2018)Mertelj, Cmok, Sebasti{\'{a}}n, Mandle,
  Parker, Whitwood, Goodby, and {\v{C}}opi{\v{c}}]{mertelj2018}
A.~Mertelj, L.~Cmok, N.~Sebasti{\'{a}}n, R.~J. Mandle, R.~R. Parker, A.~C.
  Whitwood, J.~W. Goodby and M.~{\v{C}}opi{\v{c}}, \emph{Physical Review X},
  2018, \textbf{8}, 041025\relax
\mciteBstWouldAddEndPuncttrue
\mciteSetBstMidEndSepPunct{\mcitedefaultmidpunct}
{\mcitedefaultendpunct}{\mcitedefaultseppunct}\relax
\EndOfBibitem
\bibitem[Mandle and Mertelj(2019)]{mandle2019}
R.~J. Mandle and A.~Mertelj, \emph{Phys. Chem. Chem. Phys.}, 2019, \textbf{21},
  18769--18772\relax
\mciteBstWouldAddEndPuncttrue
\mciteSetBstMidEndSepPunct{\mcitedefaultmidpunct}
{\mcitedefaultendpunct}{\mcitedefaultseppunct}\relax
\EndOfBibitem
\bibitem[Connor and Mandle(2020)]{connor2020}
P.~L.~M. Connor and R.~J. Mandle, \emph{Soft Matter}, 2020, \textbf{16},
  324--329\relax
\mciteBstWouldAddEndPuncttrue
\mciteSetBstMidEndSepPunct{\mcitedefaultmidpunct}
{\mcitedefaultendpunct}{\mcitedefaultseppunct}\relax
\EndOfBibitem
\bibitem[Sebasti{\'{a}}n \emph{et~al.}(2020)Sebasti{\'{a}}n, Cmok, Mandle,
  de~la Fuente, {Dreven{\v{s}}ek Olenik}, {\v{C}}opi{\v{c}}, and
  Mertelj]{sebastian2020}
N.~Sebasti{\'{a}}n, L.~Cmok, R.~J. Mandle, M.~R. de~la Fuente,
  I.~{Dreven{\v{s}}ek Olenik}, M.~{\v{C}}opi{\v{c}} and A.~Mertelj, \emph{Phys.
  Rev. Lett.}, 2020, \textbf{124}, 037801\relax
\mciteBstWouldAddEndPuncttrue
\mciteSetBstMidEndSepPunct{\mcitedefaultmidpunct}
{\mcitedefaultendpunct}{\mcitedefaultseppunct}\relax
\EndOfBibitem
\bibitem[{\v{C}}opi{\v{c}} and Mertelj(2020)]{copic2020}
M.~{\v{C}}opi{\v{c}} and A.~Mertelj, \emph{Phys. Rev. E}, 2020, \textbf{101},
  022704\relax
\mciteBstWouldAddEndPuncttrue
\mciteSetBstMidEndSepPunct{\mcitedefaultmidpunct}
{\mcitedefaultendpunct}{\mcitedefaultseppunct}\relax
\EndOfBibitem
\bibitem[Chaturvedi and Kamien(2019)]{chaturvedi2019}
N.~Chaturvedi and R.~D. Kamien, \emph{Phys. Rev. E}, 2019, \textbf{100},
  022704\relax
\mciteBstWouldAddEndPuncttrue
\mciteSetBstMidEndSepPunct{\mcitedefaultmidpunct}
{\mcitedefaultendpunct}{\mcitedefaultseppunct}\relax
\EndOfBibitem
\bibitem[Selinger(2018)]{selinger2018}
J.~V. Selinger, \emph{Liq. Cryst. Rev.}, 2018, \textbf{6}, 129--142\relax
\mciteBstWouldAddEndPuncttrue
\mciteSetBstMidEndSepPunct{\mcitedefaultmidpunct}
{\mcitedefaultendpunct}{\mcitedefaultseppunct}\relax
\EndOfBibitem
\bibitem[Selinger \emph{et~al.}(1993)Selinger, Wang, Bruinsma, and
  Knobler]{selinger1993}
J.~V. Selinger, Z.-G. Wang, R.~F. Bruinsma and C.~M. Knobler, \emph{Phys. Rev.
  Lett.}, 1993, \textbf{70}, 1139--1142\relax
\mciteBstWouldAddEndPuncttrue
\mciteSetBstMidEndSepPunct{\mcitedefaultmidpunct}
{\mcitedefaultendpunct}{\mcitedefaultseppunct}\relax
\EndOfBibitem
\end{mcitethebibliography}
\bibliographystyle{rsc}
\end{document}